# Advanced Risk Prediction and Stability Assessment of Banks Using Time Series Transformer Models

Wenying Sun
Southern Methodist University
Dallas, USA

Zhen Xu
Independent Researcher
Shanghai, China

Wenqing Zhang
Washington University in St. Louis
St. Louis, USA

Kunyuan Ma
New York University
New York, USA

You Wu
College of William & Mary
Williamsburg, USA

Mengfang Sun *
Stevens Institute of Technological
Hoboken, USA

*Abstract*—This paper aims to study the prediction of the bank stability index based on the Time Series Transformer model. The bank stability index is an important indicator to measure the health status and risk resistance of financial institutions. Traditional prediction methods are difficult to adapt to complex market changes because they rely on single-dimensional macroeconomic data. This paper proposes a prediction framework based on the Time Series Transformer, which uses the self-attention mechanism of the model to capture the complex temporal dependencies and nonlinear relationships in financial data. Through experiments, we compare the model with LSTM, GRU, CNN, TCN and RNN-Transformer models. The experimental results show that the Time Series Transformer model outperforms other models in both mean square error (MSE) and mean absolute error (MAE) evaluation indicators, showing strong prediction ability. This shows that the Time Series Transformer model can better handle multidimensional time series data in bank stability prediction, providing new technical approaches and solutions for financial risk management.

## I. INTRODUCTION

The bank stability index is an important indicator for measuring the health status and risk resistance of financial institutions. Especially in the context of high volatility in global financial markets, the stability of the banking system is crucial to the normal operation of national and regional economies [1]. With the rapid development of information technology and financial markets, the external risks faced by the banking industry have become more complex and diverse. How to accurately predict the stability of banks has become an important issue in financial supervision and risk management. Traditional prediction methods mainly rely on macroeconomic data and expert judgment, and often have problems such as single data dimension and inability to reflect market changes in a timely manner. In recent years, with the rapid progress of technologies such as natural language processing [2] and emotion-aware intelligence [3], financial risk prediction methods based on machine learning [4] and Convolutional Neural Networks [5] have gradually become a research hotspot. Based on the time series Transformer model [6], this paper aims to provide a new technical approach for the prediction of the bank stability index by constructing a more effective prediction framework.

The bank stability index is composed of critical financial indicators that reflect a bank's financial health and risk resistance. These include the capital adequacy ratio, a measure of a bank's available capital against risks; the non-performing loan ratio, which quantifies problematic loans; and the liquidity coverage ratio, which assesses short-term liquidity resilience. To further evaluate the importance of these features, we leveraged the attention weights in the Transformer model. For instance, the attention heatmaps reveal a dominant focus on the liquidity coverage ratio during periods of financial volatility, while long-term trends favor the capital adequacy ratio as the primary indicator. This nuanced understanding underscores the model's ability to dynamically adjust its focus based on market conditions. These indicators reflect the financial status and risk tolerance of banks from different dimensions. With the intensification of global economic uncertainty, especially the impact of events such as the COVID-19 pandemic and geopolitical conflicts, financial markets have fluctuated violently, and the pressure on the banking system has increased significantly. Traditional prediction methods based on statistical models have gradually become difficult to cope with. Statistical models usually rely on the assumption of stationarity of historical data. However, under extreme market conditions, the banking system may undergo nonlinear or even sudden changes, and traditional methods are difficult to accurately capture these complex dynamics. Based on this, the introduction of deep learning models, especially the Transformer model, provides new possibilities for the prediction of the bank stability index.

The Transformer model was proposed by Liang [7] in the field of natural language processing [8], which is the mainstream method for various sequence modeling tasks [9]. Unlike traditional models such as recurrent neural networks

(RNNs) [10] and long short-term memory networks (LSTMs) [11], the Transformer model uses a self-attention mechanism, which can not only process sequence data in parallel but also capture the dependencies between long-distance data. This feature also makes the transformer model show great potential in the field of time series prediction. The prediction of the bank stability index is a typical time series task, which contains a large amount of historical data, and these data often have complex time series dependencies and periodic fluctuation characteristics. By introducing the Transformer model, the nonlinear interactions between multiple indicators within the banking system can be better captured, thereby improving the prediction accuracy.

The Time Series Transformer excels in predicting bank stability index due to its ability to capture long-term dependencies. Traditional models often overlook cyclicality and suddenness in financial market volatility, leading to inaccurate predictions. The self-attention mechanism assigns weights to each time step, focusing on important information over longer periods. This enables improved prediction performance for complex time series data. For instance, short-term events like interest rate adjustments can have immediate impacts, while long-term trends like regulatory policy changes accumulate risks. The Transformer can simultaneously capture these two types of information. This research addresses the shortcomings of traditional models in long-term dependency modeling and multidimensional data processing, improving prediction accuracy. With the growing uncertainty of global financial market risks, an efficient bank stability prediction model is crucial for financial regulators, bank managers, and others to take early countermeasures and prevent systemic risks. Additionally, this research promotes the application of Transformer models in finance and provides new technical tools and ideas for financial data analysis.

## II. RELATED WORK

Advancements in deep learning have enabled significant improvements in predicting complex financial indices, such as the bank stability index, by addressing the limitations of traditional models in handling high-dimensional, non-linear, and temporal data.

Hybrid architectures combining convolutional and recurrent layers have demonstrated their effectiveness in capturing temporal dependencies and feature hierarchies. J. Yao et al. [12] showcased the ability of CNN-LSTM frameworks to enhance risk prediction by leveraging convolutional layers for feature extraction and LSTM layers for sequential dependency modeling. These concepts directly inform the development of hybrid solutions for complex time series prediction tasks.

The use of Transformer models, with their self-attention mechanisms, represents a key breakthrough in modeling long-term dependencies in sequential data. Y. Wei et al. [13] illustrated the superiority of Transformer-based architectures in analyzing complex datasets, a capability that this study extends to the prediction of bank stability indices by processing multidimensional temporal dependencies. Furthermore, X. Wang et al. [14] demonstrated the benefits of spatiotemporal modeling with hybrid architectures, emphasizing adaptability in processing dynamic data—a concept leveraged here to improve the robustness of financial time series predictions.

Techniques from graph neural networks (GNNs) have also provided insights into capturing relational and structural data dependencies. J. Du et al. [15] explored entity and relationship modeling in complex datasets, emphasizing the importance of understanding interactions, which aligns with this study's focus on capturing intricate multi-variable relationships in financial indices.

Recent progress in self-supervised learning offers avenues for enhancing model robustness, particularly in financial contexts where labeled data is often sparse or noisy. Y. Xiao [16] proposed methods to improve few-shot learning through self-supervised paradigms, which inspire techniques to generalize models trained on limited financial datasets to new conditions or emerging risks.

Innovations in enhancing data representation, such as those proposed by Y. Feng et al. [17] with GAN-based approaches, provide additional perspectives on improving model performance in challenging financial prediction tasks. Similarly, metric learning techniques [18] contribute methodologies for improving feature extraction and representation, which are essential for modeling high-dimensional financial indicators.

Finally, adaptive feature extraction techniques [19-20], inform this study by providing strategies to dynamically select relevant features in large-scale, multidimensional datasets. These techniques, combined with insights from machine learning applications to irregular temporal data [21], enrich the predictive framework employed in this paper.

In summary, this research builds upon these foundational contributions, integrating deep learning techniques such as self-attention, hybrid architectures, and robust data representations to design an effective framework for predicting the bank stability index. The adoption of the Time Series Transformer specifically addresses challenges related to long-term dependency modeling and multi-variable interactions, advancing the state-of-the-art in financial risk prediction.

## III. ALGORITHM PRINCIPLE

In this study, the bank stability index is predicted based on the Time Series Transformer model. The entire method framework includes steps such as data input, feature extraction, time series dependency modeling, prediction output, and loss function optimization. First, it is assumed that the bank stability data can be represented as a time series data set, denoted as $X = \{x_1, x_2, ..., x_T\}$, where $x_t \in R$ represents the d-dimensional feature vector at time t. These features can be different indicators related to bank risks. Our goal is to predict the future bank stability index $y_T \in R$ based on time series data. Its overall architecture is shown in Figure 1.

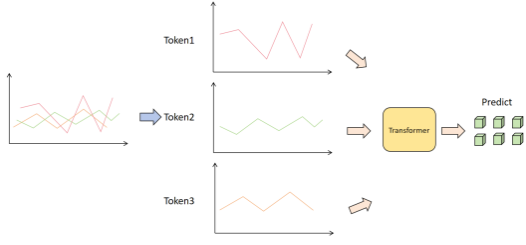

Figure 1 Overall network architecture diagram

In order to capture the dependencies between features in the time series, we first embed the input data. We first map each input feature to a high-dimensional space through a linear transformation to form an input embedding representation:

$$Z_t = W_e x_t + b_e$$

Among them, $W_e \in R^{d' \times d}$ is the embedding matrix, $b_e \in R^{d'}$ is the bias term, and $d'$ is the embedding dimension. In order to maintain the time information, Positional Embedding is used to incorporate the time information into the embedding representation:

$$PE(t, 2i) = \sin(\frac{t}{10000^{2i/d'}})$$

$$PE(t, 2i+1) = \cos(\frac{t}{10000^{2i/d'}})$$

Add the positional encoding $PE_t$ to the input embedding to get the final input representation of the time series:

$$H_t = Z_t + PE_t$$

Next, we use the multi-head self-attention mechanism [22] in Transformer to model the dependencies between time steps in the time series. For each time step t, the self-attention mechanism calculates the dependencies between this time step and other time steps. First, for each input $H_t$, we calculate the query, key, and value representation:

$$Q_t = W_q H_t$$
$$K_t = W_k H_t$$
$$V_t = W_v H_t$$

Where $W_q, W_k, W_v \in R^{d' \times d'}$ is a learnable weight matrix. Through the self-attention mechanism, we calculate the attention score at each time step:

$$a_{t,t'} = \frac{\exp(\frac{Q_t \cdot K^T}{\sqrt{d'}})}{\sum_{t'=1}^{T} \exp(\frac{Q_t \cdot K^T}{\sqrt{d'}})}$$

The attention score $a_{t,t'}$ reflects the strength of the dependence of time step t on other time steps t'. Then, these attention weights are used to weight the sum of the value vectors of all time steps to obtain the output representation of time step t:

$$O_t = \sum_{t'=1}^{T} \alpha_{t,t'} V_{t'}$$

In order to enhance the expressiveness of the model, a multi-head attention mechanism is used to perform parallel calculations on different attention heads and then concatenate their results:

$$O_{multi-head} = Concat(O_{t1}, O_{t2}, ..., O_{tn}) W_o$$

After the multi-head self-attention output is layer normalized and feed-forward neural network (FFN), the final time-dependent feature representation is obtained:

$$H_t^{output} = FFN(LayerNorm(O_{multi-head}))$$

After obtaining the feature representation of each time step, we use a fully connected layer to map the feature representation of the last time step T to the prediction result [23], that is, the predicted value of the bank stability index:

$$y_T = W_y H_T^{output} + b_y$$

Among them, $Wy \in R^{1 \times d'}$ is the learnable weight matrix and $b_y \in R$ is the bias term. In order to optimize the model parameters, we define a loss function based on the mean squared error (MSE) to measure the gap between the predicted value and the true value:

$$L = \frac{1}{N} \sum_{i=1}^{N} (y_T^{(i)} - y_T'^{(i)})^2$$

Where N is the number of samples, $y_T^{(i)}$ and $y_T'^{(i)}$ represent the true value and predicted value of the i-th sample, respectively. By minimizing the loss function, the model can gradually learn the time series dynamic changes of the bank stability index, thereby improving the prediction accuracy.

The training process of the entire model is carried out through backpropagation and gradient descent, and the optimization goal is to minimize the loss function L. Through continuous iteration, the model can learn the potential time series dependence and risk characteristics of the banking system from historical data, so as to make more accurate stability predictions at future time points.

IV. EXPERIMENT

A. Datasets

In order to conduct the bank stability index prediction experiment, this paper selected the Bank Marketing Dataset, which comes from a bank in Portugal and is widely used to study bank marketing and financial risk-related issues. The dataset contains more than 45,000 records, including detailed information on bank customers, including 16 features. These features include the customer's age, occupation, marital status, educational background, loan history, bank balance, and interaction information with the bank (such as whether to participate in a fixed deposit, the number of past contacts, the duration of the last contact, etc.). The target variable is whether the customer has subscribed to a fixed deposit plan, which is one of the important measures of bank stability and profitability. Although this dataset is mainly used for marketing strategy analysis, by further processing these features, we can use it for the task of bank risk prediction. For example, the customer's loan status, economic background, and interaction behavior with the bank may have an impact on the overall risk stability of the bank. We can use these features as input variables and use the Time Series Transformer model to predict the future risk and stability changes of the bank, thereby helping the bank to conduct more effective risk management and decision-making.

### B. Experimental Results

We compared the bank stability index prediction effect of the Time Series Transformer model with five commonly used deep learning models: LSTM, GRU, CNN, TCN, and RNN-Transformer. LSTM has good memory and captures long-term dependencies, while GRU is faster and retains important information. CNN extracts spatial features, and TCN combines RNN and CNN for long-term dependencies and parallelization. RNN-Transformer combines RNN and Transformer for sequential information and long-distance dependencies. By comparing these models, we can evaluate the Time Series Transformer's effectiveness in bank stability prediction.

Table 1 Experimental results

| Model | MSE | MAE |
|---|---|---|
| LSTM | 0.0423 | 0.1675 |
| GRU | 0.0391 | 0.1582 |
| CNN | 0.0364 | 0.1471 |
| TCN | 0.0337 | 0.1358 |
| RNN-Transformer | 0.0305 | 0.1226 |
| Ours | 0.0271 | 0.1120 |

From the experimental results in Table 1, the model based on Time Series Transformer (Ours) performs better than other models in the task of predicting the bank stability index. By comparing the two evaluation indicators of MSE and MAE, it can be found that the traditional LSTM and GRU models have certain limitations in capturing long-term dependencies in time series. The MSE of LSTM is 0.0423 and the MAE is 0.1675. Although GRU has simplified its structure and improved its computational efficiency, its MSE and MAE are 0.0391 and 0.1582 respectively, with limited improvement. This shows that when dealing with the time dependency of complex financial data, the performance of LSTM and GRU is relatively limited, especially when faced with long-term dependencies and nonlinear relationships, they cannot fully exert their potential, resulting in high prediction errors.

In contrast, the performance of CNN and TCN models has improved. CNN effectively extracts local features through convolution operations, and can capture short-term fluctuation patterns when processing time series data, so MSE and MAE are reduced to 0.0364 and 0.1471 respectively. However, due to the lack of CNN in modeling the global dependency of time series, it only relies on local feature extraction and fails to effectively capture long-term trend changes. The TCN model introduces the mechanism of temporal convolution, combines the advantages of convolutional networks and recurrent networks, can better handle long-term and short-term dependencies, and further improves the prediction accuracy. The MSE is reduced to 0.0337 and the MAE is 0.1358. This shows that TCN has a good overall performance in processing financial time series tasks, especially in parallel computing and long-term dependency modeling.

Finally, the RNN-Transformer and Time Series Transformer models show the best performance, especially the model based on the Transformer architecture. RNN-Transformer combines the sequential information processing ability of RNN and the self-attention mechanism of Transformer, which can better capture long-distance dependencies. Its MSE and MAE are 0.0305 and 0.1226 respectively, which are significantly better than the previous models. The Time Series Transformer model performs well, with the MSE further reduced to 0.0271 and the MAE reaching 0.1120, becoming the best model. The self-attention mechanism of Transformer can not only process dependencies in time series in parallel, but also effectively model the relationship between complex financial features by focusing on the global dependencies between different time steps. This shows that the Time Series Transformer model can better capture the dynamic changes of multi-dimensional financial data when dealing with the task of predicting the bank stability index, and improve the prediction accuracy and stability of the model. Finally, we show how MAE and MSE continue to decrease during training.

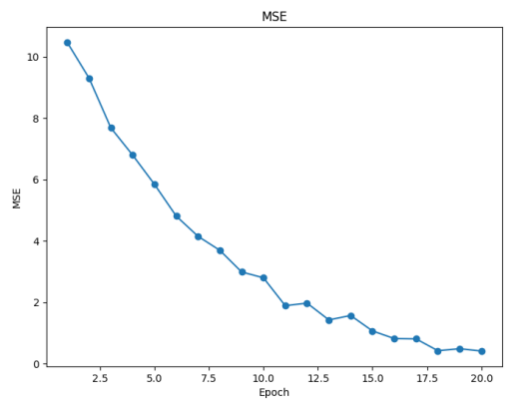

Figure 2 MSE training process decline graph

As shown in Figures 2 and 3, MSE and MAE decrease rapidly in the first 12 epochs, and then gradually converge and fluctuate, which also shows that our model can converge quickly.

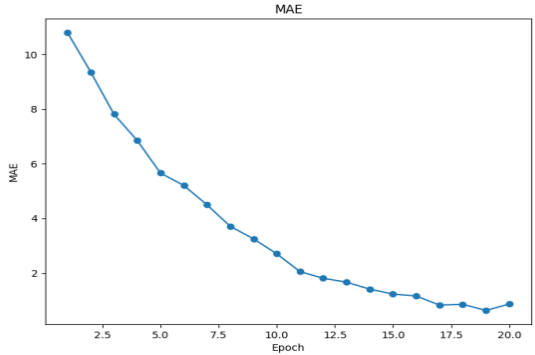

Figure 3 MSE training process decline graph

## V. CONCLUSION

This paper predicts the bank stability index based on the Time Series Transformer model, and verifies the effectiveness of this method through comparative experiments with five other deep learning models. It can be seen from the experimental results that traditional recurrent neural network models (LSTM, GRU) have certain limitations in capturing long-term dependencies in time series, resulting in their relatively poor performance in processing complex financial data. Although the CNN and TCN models perform well in local feature extraction, they still have shortcomings in long-term dependency modeling. In contrast, models based on the Transformer architecture (RNN-Transformer and Time Series Transformer) better capture the complex global dependencies in time series through the self-attention mechanism, showing significant performance improvements. In particular, the Time Series Transformer model performs best in both MSE and MAE indicators, verifying its advantages in processing multi-dimensional time series data. The proposed model offers actionable insights for financial regulators and bank managers by providing timely and accurate stability predictions. Regulators can utilize these predictions to prioritize interventions, enforce compliance, and avert systemic risks. For banks, the model aids in optimizing risk portfolios and planning for potential liquidity crises. However, deploying this model in real-world scenarios entails challenges such as integrating disparate data sources, ensuring model interpretability for stakeholders, and addressing computational scalability for large-scale implementation. Furthermore, regulatory compliance regarding data usage and potential biases in predictions must be carefully managed to ensure equitable and reliable outcomes.